\documentstyle[aps,prl,twocolumn,epsf,floats]{revtex}

\begin{document}

\draft
\twocolumn[\hsize\textwidth\columnwidth\hsize\csname
@twocolumnfalse\endcsname

\title{Adsorbate-induced substrate relaxation and 
the adsorbate--adsorbate interaction}
\author{R. Brako and D. \v{S}ok\v{c}evi\'{c}}
\address{ Rudjer Bo\v{s}kovi\'{c} Institute, P.O.~Box 1016, 10001 Zagreb, 
Croatia}%
\date{\today}
\maketitle
\thispagestyle{empty}
\widetext
\begin{abstract} 
We formulate the theory of the
perturbation caused by an adsorbate upon the substrate 
lattice in terms of a local modification of the interatomic potential
energy around the adsorption site, which leads to the relaxation of
substrate atoms. We apply the approach to
CO chemisorption on close-packed metal surfaces, and show that 
the adsorbate--adsorbate interaction and a variety of other
properties can be well described by a simple model.
\end{abstract}
]

\narrowtext

Several direct and indirect (through-the-substrate) mechanisms
can lead to an effective interaction between adsorbates on metal surfaces.
The possible role of adsorbate-induced substrate relaxation 
was considered
some time ago by Lau and Kohn \cite{Lau77,Lau78}, using an elastic
continuum model of the surface.
They concluded that the resulting interaction was 
repulsive between identical adsorbates, varied as $\rho^{-3}$ with
separation $\rho$, and was inversely proportional to the
shear modulus of the substrate. 
For adsorbates separated by a few atomic spacings, however, the
continuum theory may not be valid.
More recently, in a series of papers Kevan 
et al.\ \cite{Kevan98,Skelton94,Skelton97,Wei95,Wei97} 
have determined adsorbate--adsorbate interaction
energies for CO at several metal surfaces, using a 
transfer-matrix analysis of thermal desorption spectra.
They have also  qualitatively discussed the adsorbate-induced 
strain as a possible
mechanism of the adsorbate--adsorbate interaction at intermediate
range, i.e., several substrate atoms apart.

The potential energy of the atomic lattice of a solid in
the harmonic approximation can be written as
\begin{equation}
\label{harmonic}
V = \sum_{i,j,\mu, \nu}
x_\mu (i) D_{\mu \nu} (i,j) x_\nu (j) ,
\end{equation}
where $x_\mu (i)$ is the $\mu$-th component of the displacement
of the $i$-th atom from the equilibrium position.
Now assume that an impurity is created by replacing one atom with 
a different species. (In the general discussion we talk about an
``impurity'', although we are primarily interested 
in the chemisorption case, where the adsorbate also introduces 
additional degrees of freedom. The generalization to the latter
case is straightforward.) The potential energy after the replacement 
can again be expressed in the form~(\ref{harmonic}), but the new 
equlibrium positions $x'_\mu (i)$ are in general different:
\begin{equation}
\label{impurity}
x'_\mu (i) = x_\mu (i) + \Delta x_\mu (i) .
\end{equation}
The new dynamical matrix $D'_{\mu \nu} (i,j)$ is also different, and a 
constant term appears which shifts the energy minimum.
Using~(\ref{impurity}), 
the potential energy of the system after the impurity has
been introduced can be expressed in
the coordinates $x_\mu (i)$:
\begin{equation}
\label{harmonic2}
V' = \sum_{i,j,\mu, \nu}
x_\mu (i) D'_{\mu \nu} (i,j) x_\nu (j) 
+ \sum_{i,\mu} F_\mu (i) x_\mu (i) + V'_0,
\end{equation}
i.e., the dynamical matrix is modified, and linear (force)
and constant (energy shift) terms appear.

The foregoing considerations depend only upon the assumed 
stability of the solid, i.e., the existence of the minimum of the
potential energy. 
For our application to
chemisorption, we specifically assume the following properties:
(a)~the force terms $F_\mu (i)$ are nonzero only for a small number 
of substrate atoms around the adsorption site; (b)~similarly, only a few 
elements of the dynamical matrix $D_{\mu \nu} (i,j)$ change, if any;
(c)~the effect is linear, so that the chemisorption of another molecule
(and, consequently, of a third, a fourth, etc.) can be described by the same
set of parameters, of course centered around the 
new adsorption site. The condition~(c) is rather restrictive
and excludes systems which reconstruct at large adsorbate coverage, 
as well as those where other interactions are important, such as the
direct adsorbate--adsorbate repulsion, the electrostatic dipole--dipole
interaction, or the ``chemical'' competition for the same electronic
orbitals in the substrate.
Estimates, however, show that such interactions are
usually weak and short-ranged, 
while we are interested in medium-range interactions (second
nearest neighbor and beyond) of nonionic adsorbates.

In this paper we take into account the in-plane relaxation within the first 
atomic layer of the substrate induced by a chemisorbed species. 
The relaxation in the perpendicular direction can, of course, be 
equally strong, but it does not contribute much to the effective
adsorbate--adsorbate interaction and to the adsorbate-induced
surface stress.
Also, we do not consider the coupling to internal adsorbate coordinates,
which was discussed in Ref.~\cite{Brako98} in an application to the
damping of adsorbate vibrations.

For a hexagonal layer of atoms, we write the potential energy as
\begin{eqnarray}
V & = & \frac{1}{2}
 \sum_{i} \sum_{j=1}^{6} \frac{1}{2}K_1 [\hat{{\bf{r}}}_{ij} \cdot 
 ({\bf{r}}_i - {\bf{r}}_j )]^2 + \frac{1}{2} \sum_{i}K_2 {\bf{r}}_i^2,
\label{potentialsubstrate}
\end{eqnarray}
where $\bf{r}_i=(x_i,y_i)$ is the in-plane displacement from the
equlibrium position of the $i$-th atom. The term $K_1$ describes
a central atom--atom interaction, and the term $K_2$ binds atoms
to their equilibrium positions, simulating the interaction to lower
atomic layers. Without it, the model would be too ``soft'' to
long-wavelength perturbations. The trade-off is that the lowest
phonon frequency becomes finite, i.e., there are no true ``acoustic'' 
modes, but this has little influence on the relaxation energy
and other static quantities calculated in this work.

Now assume that an atom or a molecule chemisorbs on top of the atom 
$i = 0$, and that the induced change of the potential energy is
\begin{eqnarray}
V' &=& \sum_{j=1}^{6} \frac{1}{2} 
\Delta K_1 [\hat{{\bf{r}}}_{0j} \cdot
 ({\bf{r}}_0 - {\bf{r}}_j )]^2 \nonumber \\
  && - \sum_{j=1}^6\frac{k_2}{\alpha} \hat{\bf{r}}_{0j}\cdot ({\bf{r}}_0 -
 {\bf{r}}_j) + V'_0, 
\label{potentialinteraction}
\end{eqnarray}
where the first term describes the change of the force constant between 
the atom $0$ 
and the six surrounding atoms, and the second 
is a linear force term.
In the following we drop the $\Delta K_1$ and $V'_0$ terms, which are 
beyond the level of accuracy assumed in this work, 
although, in general, the $\Delta K_1$ term ought to be included.
 
The definition of the chemisorption-induced surface stress $\tau$ 
is~\cite{Ibach97}
\begin{equation}
\delta W = A \tau_x \delta\epsilon_x,
\end{equation}
where $A$ is the surface area, 
$\delta\epsilon_x$ the strain, and $\delta W$ the difference of the
work involved in straining a clean surface 
and a surface with adsorbates.
In our model, we obtain
\begin{equation}
\tau_x =- \frac{k_2}{a \alpha} 2 \sqrt{3}\ \theta,
\label{stressontop}
\end{equation}
where $\theta$ is the adsorbate coverage.
(If $ \Delta K_1$ is different from zero, an additional factor of 
order unity appears
in the above expression.)

\begin{figure}[ht]
\centerline{\epsfxsize=0.20\textwidth
\epsfbox{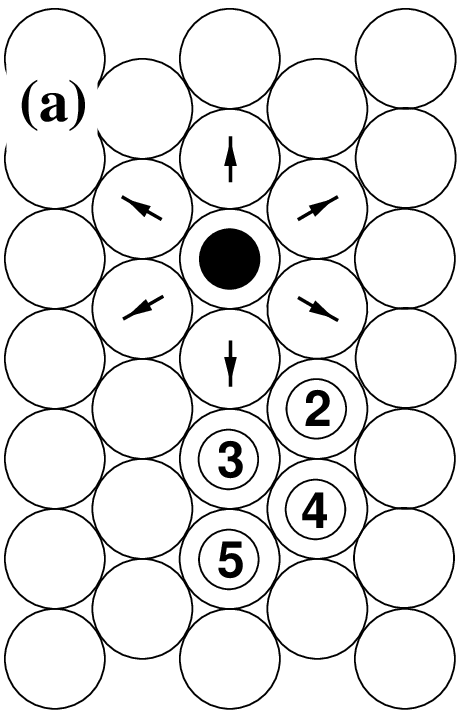} \hspace*{0.2cm}
\epsfxsize=0.20\textwidth
\epsfbox{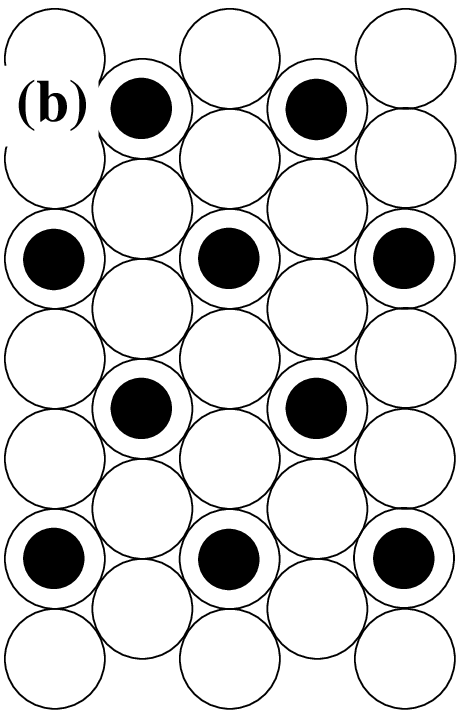}}
\vspace{0.2cm}
\caption{
(a) Chemisorption of CO, black circle, into an on-top site on
Pt(111). The chemisorbate causes a relaxation of the
adjacent Pt atoms, denoted by arrows, which propagates to further first-layer
atoms via elastic forces. 
The neighboring adsorption sites are 
indicated by numbers.
(b) The regular $(\sqrt{3}\times\sqrt{3}){\rm R}30^{\circ} $
structure at a coverage $ \theta = 0.33$. Owing to symmetry, no relaxation
of substrate atoms along the surface plane is possible.
   }
\label{fig:Pt}
\end{figure}

Irrespective of the sign of the force term $k_2/\alpha$, there is
always an energy gain due to the relaxation of the surrounding atoms, 
i.e., the minimum of $V + V'$ is less than zero. 
If another molecule adsorbs on a nearby site,
the forces induced by the two adsorbates
act in opposite directions and the relaxation is
less complete than with adsorbates far apart,
which leads to an effective interaction. We calculate
the interaction energies by comparing
the minimum of the potential energy $V + V'$ for a single
adsorbed molecule with the minimum for two molecules adsorbed on,
in turn, second, third, fourth, and fifth nearest-neighbor
sites, see Fig.~\ref{fig:Pt}(a). 
(Throughout the paper we assume that there is a strong repulsion
between first nearest neighbor adsorbates caused by ``chemical'' 
effects, and do not consider them.)

We first consider the chemisorption of CO on a Pt(111) surface.
There is a large amount of tensile stress  in the first atomic layer of the
clean Pt(111) surface~\cite{Hohage95,Boisvert97}. Although the
surface does not
reconstruct at room temperature, a reconstruction is observed at high
temperatures in the presence of saturated Pt vapor~\cite{Hohage95}.
CO adsorbs initially into on-top sites of Pt(111), but the
energy difference for the adsorption into bridge sites is
obviously small, since some bridge adsorbates are
found already at coverages $\theta$ above 0.15 \cite{Yoshinobu96}. 
Some  authors report (Ref.~\cite{Yeo97-King} and references therein)
that a regular 
$(\sqrt{3} \times \sqrt{3}){\rm R}30^{\circ}$
structure at a coverage $\theta = 1/3$
is formed, Fig. \ref{fig:Pt}(b), but others claim that the densest
structure of exclusively on-top adsorbates  exists at $\theta = 0.29$
\cite{Tushaus87}, and that further chemisorption occurs into
bridge sites.
The regular structure at $\theta = 0.5$ contains
an equal number of on-top and bridge adsorbates.

We have chosen the values of the parameters $K_1$, $K_2$, 
and $k_2/\alpha$ which give good agreement with
experimental data on adsorbate-induced surface stress \cite{Ibach97} and
with low-coverage interaction energies 
\cite{Skelton94},
as shown in Table~\ref{tab:energies}.
We had to choose a 
rather small value for the force constant $K_1$ between atoms in the 
first surface layer. (The value of $K_2$ has little effect on the
results. 
The much larger value of $K_1$ used in a similar model in
Ref.~\cite{Brako98} was an overestimate.)
The reduction from bulk values is characteristic of
many close-packed noble-metal surfaces~\cite{Bortolani90},
but the reduction we find is larger than suggested 
in the surface phonon calculation in 
Ref.~\cite{Bortolani89}. Consequently, the highest resulting vibrational 
frequency of two-dimensional phonons of the first surface layer is
about 5 meV, by about a factor of two
smaller than the frequency of surface 
phonons along the edge of the Brillouin zone of around 10 meV
calculated in Ref.~\cite{Bortolani89}. 
As discussed above, it is possible that part of the softening
is localized around the adsorption site only, the term $\Delta K_1$
in Eq.~(\ref{potentialinteraction}). 
\begin{table}[ht]
\caption{Interaction energies $W_{n{\rm NN}}$ (in K)
and the induced surface stress $\tau$ (in N/m)
for on-top adsorption of CO on Pt(111).
 $W_{n{\rm NN}}=2 E_0-E_{n{\rm NN}}$,
  where $E_0$ is the relaxation energy of a single CO molecule,
 $E_{n{\rm NN}}$ is the relaxation energy for two CO molecules
  adsorbed at the $n$-th nearest-neighbor sites, etc.
Theoretical values are calculated using
 $K_1 = 4$~N/m, $K_2 = 2.5$~N/m,
$k_2/\alpha = 0.3 \times 10^{-9} $~N.
The nearest-neighbor distance is $a = 2.76$~\AA.
\label{tab:energies}}
\vspace{0.3cm}
\begin{tabular}{lclllll}
 & $W_{2\rm NN}$ & $W_{\rm 3NN}$ & $W_{\rm 4NN}$
                     & $W_{\rm 5NN}$ & $E_0/6 $ & $\tau$ \\ \tableline
Theory  & 194 & 314 & 108 & 119 & 231 & 1.24  \\
Experiment  & 120\tablenotemark[1]
  & 400\tablenotemark[1] & 236\tablenotemark[1]
  & & 452\tablenotemark[2] & 1.2\tablenotemark[3]
\end{tabular}
\tablenotetext[1]{\footnotesize Ref. \cite{Skelton94}, low CO coverage.}
\tablenotetext[2]{\footnotesize Ref. \cite{Yeo97-King}, CO coverage of 0.33.}
\tablenotetext[3]{\footnotesize Ref. \cite{Ibach97}, CO coverage of 0.33.}
\end{table}
The values in Table~\ref{tab:energies} show that the repulsive
interaction is strong between CO adsorbed on sites lying
along rows of substrate atoms, and weaker for adsorbates 
separated by hollows, even if they are less far apart.
In our opinion, 
the rather large
interaction energy between fourth nearest neighbor adsorbates
in Ref.~\cite{Skelton94} is influenced
by the contributions from more distant sites, which were not
included in their analysis.

\begin{figure}[ht]
\centerline{\epsfxsize=0.20\textwidth
\epsfbox{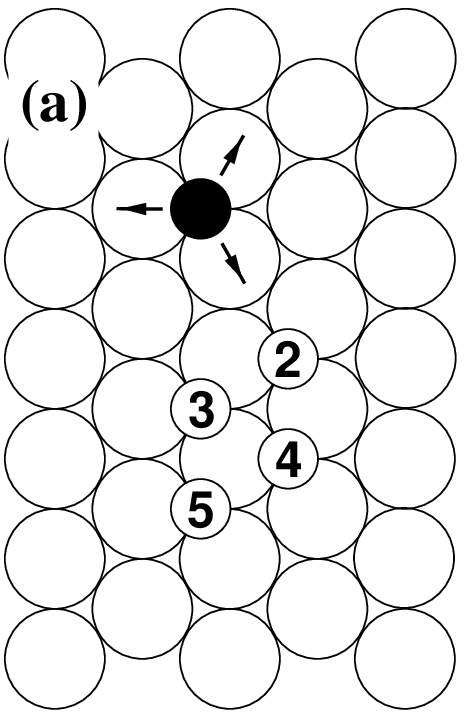} \hspace*{0.2cm}
\epsfxsize=0.20\textwidth
\epsfbox{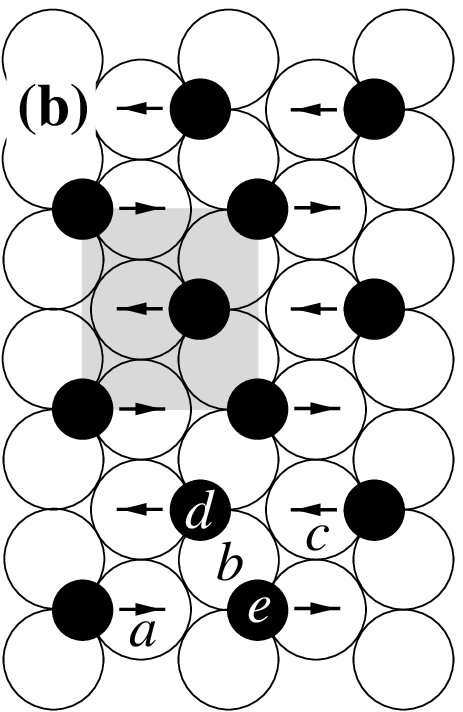}}
\vspace{0.2cm}
\caption{ (a) Chemisorption of CO, black circle, into a three-fold
hollow site on Ni(111). The relaxation of the
adjacent Ni atoms is denoted by arrows.
The numbers indicate 
possible adsorption 
sites of the same kind, either fcc or hcp.
(b) The regular c$(4 \times 4)$--2CO
structure on Ni(111) at a coverage $\theta = 0.5$. An elementary cell
(grey area) contains one adsorbate in the fcc and one in the hcp position,
{\it d} and {\it e}. The relaxation of Ni atoms in intermediate rows,
{\it a} and {\it c}, is denoted by arrows.
  }
\label{fig:Ni}
\end{figure}

Clean nickel 
(111) surfaces do not reconstruct.
Unlike some earlier claims, it is now accepted that at low temperature CO 
chemisorbs initially into threefold hollow sites \cite{Chen89,Held98}. 
At room temperature, some bridge and on-top sites seem to be occupied
even at low coverages \cite{Held98}.
At a coverage $ \theta = 0.33$, CO forms a
regular $(\sqrt{3}\times\sqrt{3}){\rm R30}^{\circ}$ structure~\cite{Becker93},
but it is not clear whether the molecules adsorb into
fcc or hcp positions. At $\theta = 0.5$, a regular c(4$\times$2)--2CO
structure is formed, in which an equal number of fcc and hcp sites is
occupied \cite{Becker93,Mapledoram94,Sprunger95}, Fig.~\ref{fig:Ni}(b).
The top layer of Ni atoms shows buckling in that nonequivalent atoms have
different vertical relaxation. 
The CO molecules are slightly tilted from the direction perpendicular
to the surface.
The initial heat of adsorption at room temparature is 
130 kJ/mol~\cite{Stuckless93}. It decreases slowly at
first, to around 122 kJ/mol at $\theta = 0.33$ and 112 kJ/mol at 
$\theta = 0.5$. 
However, the fact that not all adsorption sites are equivalent,
and the rather large standard deviation of experimental data make
the interpretation of the coverage dependence of the heat of adsorption
uncertain.

We describe the first layer of Ni atoms by the same potential as for Pt,
Eq.~(\ref{potentialsubstrate}).
The interaction terms are similar to Eq.~(\ref{potentialinteraction}),
but the adsorbate is in a threefold hollow site and the sums run over
the three surrounding Ni atoms (Fig.~\ref{fig:Ni}).
The induced surface stress is
\begin{equation}
\tau = - \frac{k_2}{\alpha a} \theta,
\label{stresshollow}
\end{equation}
where 
the nearest-neighbor Ni--Ni distance is  $a = 2.49$~\AA. 
From the experimental data $\tau = -0.55$
N/m at $\theta = 0.33$~\cite{Grossmann94} we have estimated 
$k_2/\alpha = 1 \times 10^{-10}$~N.
In a LEED analysis of the c$(4 \times 2)$--2CO structure which forms
at  $\theta = 0.5$, Mapledoram et al.~\cite{Mapledoram94} found
that the lateral displacement of the Ni atoms next to an adsorbate,
{\it a} and {\it c} in Fig.~\ref{fig:Ni}(b), was 0.03~\AA. 
We have reproduced this value by taking $K_1 = 6$~N/m and $K_2 = 2$~N/m.
(As before, the value of $K_2$ is of lesser importance.)
This is a considerable reduction from the bulk values, 
but not as large as 
for the Pt(111) surface, in agreement with the fact that the Ni(111)
surface seems less prone to reconstruct than Pt(111). The energy gain
per adsorbate is 22~K. 

Using these values of the parameters,
the relaxation energy for a single adsorbate is only 42~K, 
which is 30 times smaller than that
we have found for CO/Pt(111). (This reduction can be well estimated treating
each surrounding Ni atom as an independent oscillator: The force 
$k_2/\alpha$ is
three times smaller than in the CO/Pt(111) case, the Ni--Ni force
constant is 50\% larger, and there are 3 surrounding atoms instead of 6.) 
Furthermore, the displacements of Ni atoms around the hollow adsorption 
site are not along chains of atoms, and do not cause a large displacement
of other atoms.
The calculated effective interaction energies for second nearest
neighbors and beyond are therefore only a few K, too small
to be observable. 
In our opinion, the interaction 
energy of 100~K between second neighbor adsorbates suggested by 
Skelton et al.~\cite{Skelton97} is either a combined result of
other mechanisms or an artefact of their procedure. We note that
an earlier study~\cite{Gijzeman84}
reported that there was essentially no interaction already between
second neighbor adsorbates. 

Interaction energies have also been determined for CO chemisorbed 
on Rh~\cite{Wei97,Payne92} and Cu~\cite{Wei95} surfaces. We
discuss these systems only qualitatively, since the proposed values
are still uncertain, and there is no quantitative data on
other adsorbate-induced properties.
Wei et al.~\cite{Wei97} estimated that $W_2 = -100$~K and 
$W_3 = 150$~K for the on-top chemisorbed CO on Rh(111). An earlier
measurement by Payne et al.~\cite{Payne92} reported $W_2 = 170$~K
and $W_3 = -85$~K. In our model, the relative magnitudes of
interaction energies for on-top adsorbates on fcc (111) surfaces
are always similar to those found for CO/Pt(111). In particular,
we expect large repulsion between third nearest neighbor adsorbates
which lie along a chain of substrate atoms. In this respect the values
proposed in Ref.~\cite{Wei97} seem more probable, although the origin of the 
attractive $W_2$ (if it is real) is not clear. For the on-top CO on 
Cu(111), the same authors found 
$W_2 = 107$~K, $W_3 > 800$~K, $W_4 = 155$~K~\cite{Wei95}. 
The value of $W_3$ seems too large compared with the other two, 
but otherwise the results are quite similar to 
CO/Pt(111). 

The square lattice of the first layer of the (100) surface of the fcc bulk
is not well described by a purely pairwise potential similar to
Eq.~(\ref{potentialsubstrate}), since in the absence of 
angular force constants and 
of coupling to 
lower layers, only the {\it ad hoc} term $K_2$ ensures the stability.
Nevertheless, the trends in the interaction energies between on-top
adsorbates can be deduced by analogy with (111) surfaces. We expect 
a repulsive interaction between third nearest neighbor adsorbates
which lie on the same chain of atoms, and no interaction between
second nearest neighbors which lie diagonally on different chains,
because the forces on adjacent substrate atoms are orthogonal. In 
the latter case, even a weak attraction is possible owing to partly 
collinear displacements induced by the two adsorbates on more 
distant surface atoms. Indeed,
the values  $W_2 = 0$ and $W_3 = 400$~K were found for
Rh(100)~\cite{Wei97}, while the values suggested for Cu(100) were
$W_2 = -33$~K and $W_3 = 13$~K~\cite{Wei95}. It is interesting that 
the continuum elastic theory~\cite{Lau78} also gives a strong
repulsion in the $\langle 110 \rangle$ 
direction and possibly a weak attraction in the 
$\langle 100 \rangle$ direction between adsorbates on (100)
surfaces of noble metals.

We have shown that the adsorbate-induced substrate relaxation leads to
an adsorbate--adsorbate interaction which is quite long-ranged and has a
nonmonotonic distance dependence, being particularly large between 
molecules adsorbed in on-top positions along a chain of surface atoms. 
The same mechanism also leads to other 
observable effects,
such as the adsorbate-induced surface
stress and the relaxation displacement of substrate atoms. Quantitative
agreement with experiment can be obtained 
for CO adsorbed on Pt(111) and some other surfaces
using a simple model, 
assuming that force constants between atoms in the first surface
layer are considerably weaker than in the bulk, which is a known
property of many close-packed  noble metal surfaces. The results 
emphasize the importance of allowing the full substrate relaxation in
the first-principle calculations of chemisorption on metal surfaces.

This work was supported by the Ministry of Science and Technology
of the Republic of Croatia under contract No. 00980101.

\end{document}